\documentclass[aps,showpacs,preprintnumbers,amsmath,amssymb]{revtex4}
\usepackage{dcolumn}
\usepackage[dvips]{epsfig}

\usepackage{float}
\usepackage{graphics,epsfig}
\usepackage{graphicx}
\usepackage{dcolumn}
\usepackage{bm}
\usepackage{gensymb}
\usepackage{subfigure}
\usepackage{diagbox}
\usepackage{enumerate}
\usepackage{color}
\usepackage{amsmath}
\usepackage{ulem}
\usepackage{color}

\begin{document}
\baselineskip=0.5 cm
\title{\bf The ring-shaped shadow of rotating naked singularity with a complete photon sphere}

\author{ Mingzhi Wang$^{1}$\footnote{wmz9085@126.com}, Guanghai Guo$^{1}$, Pengfei Yan$^{1}$, Songbai Chen$^{2,3}$, Jiliang Jing$^{2,3}$}
\affiliation{$ ^1$School of Mathematics and Physics, Qingdao University of Science and Technology, Qingdao, Shandong 266061, People's Republic of China \\
$ ^2$Institute of Physics and Department of Physics, Key Laboratory of Low Dimensional Quantum Structures
and Quantum Control of Ministry of Education, Synergetic Innovation Center for Quantum Effects and Applications,
Hunan Normal University, Changsha, Hunan 410081, People's Republic of China\\
$ ^3$Center for Gravitation and Cosmology, College of Physical Science and Technology,
Yangzhou University, Yangzhou 225009, China}

\begin{abstract}
\baselineskip=0.4 cm

{\bf Abstract} We investigate the shadows of Konoplya-Zhidenko naked singularity. In the spacetime of Konoplya-Zhidenko naked singularity, not only can unstable retrograde light ring (LR) exist, but also unstable prograde LR, leading to the formation of a complete photon sphere (PS). Due to the absence of an event horizon, a dark disc-shaped shadow does not appear; instead, a ring-shaped shadow is observed. The ring-shaped shadow appears as an infinite number of relativistic Einstein rings in the image of the naked singularity. For some parameter values, only the unstable retrograde LR exists, resulting in an incomplete unstable PS and consequently giving rise to the arc-shaped shadow for Konoplya-Zhidenko naked singularity. The shadow of Konoplya-Zhidenko naked singularity gradually shifts to the right as the rotation parameter $a$ increases, and gradually becomes smaller as the deformation parameter $|\eta|$ increases. Moreover, the stable LRs and stable photon spherical orbits can also exist in Konoplya-Zhidenko naked singularity spacetime, but they have no effect on the image of the naked singularity. This study demonstrates that rotating naked singularity can exhibit not only an arc-shaped shadow but also a ring-shaped shadow.

\end{abstract}

 \maketitle
\newpage
\section{Introduction}

The Event Horizon Telescope (EHT) Collaboration et al have announced the images of supermassive black holes located at the center of the giant elliptical galaxy M87\cite{fbhs1,fbhs2,fbhs3,fbhs4,fbhs5,fbhs6} and the Milky Way Galaxy\cite{sga1,sga2,sga3,sga4,sga5,sga6}. The brightness depression inside the bright asymmetric ring in black hole image resolved by EHT is highly likely to be a black hole shadow, which is caused by the absorption of lights into event horizon\cite{synge,sha2,lumi,sha3}. The black hole shadow contains valuable information about the compact object, making it a vital tool in the study of black holes for constraining black hole parameters\cite{sha9,sha8,dressed,Intcur,bhparam,obsdep,constr}, exploring fundamental physics issues like dark matter\cite{polar7, drk, polar8,shadefl,shasgra}, and testing various gravity theories\cite{safeg,lf,sha10,fR, 2101, 2107, 2111, Jing, Jing1, Jing2023, Jing2021, Fang, Chen, Pan}. Black hole shadows have also been investigated in previous studies\cite{fpos2,sy,sb10,sw,swo,astro,chaotic,my,sMN,swo7,mbw,mgw,scc,sha16,shan1add,shan3add,pe,halo,review,lf2,Zeng2020vsj,Zeng2020dco,knn,BI,qx1,zzl1,qx2,lxy1,zzl2,qx3,bir,qx4,sdsph,sdnt,sdwps,litra,dkerr,phoreg}.

Gravitational collapse can result in the emergence of spacetime singularity. But the cosmic censorship conjecture demands that singularities should be hidden by event horizons. However, the cosmic censorship conjecture requires further verification. So the observations of naked singularities play a crucial role in testing the cosmic censorship conjecture and are also significant for general relativity. For a black hole, the appearance of a shadow is due to the light rays entering the event horizon, but its boundary is determined by the unstable PS. The PS is a region in space where photons can orbit the black hole in spherical orbits. It consists of both prograde and retrograde photon spherical orbits (PSOs). The prograde and retrograde LRs, representing the circular photon orbits in the equatorial plane, are the leftmost and rightmost orbits within the PS. If the PS is composed of a continuum of PSOs that connect the prograde and retrograde LRs, then this PS is complete. In general, naked singularity cannot cast a shadow. However, numerous studies indicate that naked singularity also can cast a shadow\cite{sdsph,sdnt,sdwps}. For some spherically symmetric naked singularities, the shadows are casted by the naked singularities themselves\cite{sdnt,sdwps}. For rotating naked singularity, unstable prograde LR is typically absent, resulting in an incomplete unstable PS, which gives rise to the appearance of an arc-shaped shadow\cite{sha9,litra}. Due to the Kerr naked ring singularity, the light rays can pass through the inside of the singular ring(the other world $r<0$), leading to a dark spot emerging in the image of Kerr naked singularity\cite{sha9}. When observed on the equatorial plane, a black straight line will emerge in the image of Kerr naked singularity due to the light rays hit on the ring singularity\cite{sha9}.

In this paper, we mainly research the LRs, PSs, and images of Konoplya-Zhidenko naked singularity. The Konoplya-Zhidenko metric describes an asymptotically flat, stationary, and axisymmetric spacetime beyond General Relativity by adding a static deformation from the Kerr spacetime, which sharply modifies the structures of spacetime in the strong-field region\cite{kz}. The effects of the deformation parameter on the quasinormal modes and superradiance of Konoplya-Zhidenko black hole have investigated in Ref.\cite{suprr,quasin}. PS and LR are closely related to the ringdown stage and the shadow of black hole. In Konoplya-Zhidenko naked singularity spacetime, both prograde and retrograde unstable LRs could exist, leading to the presence of a complete unstable PS. Furthermore, both prograde and retrograde stable LRs could also exist in this spacetime, so does the complete stable PS. This will lead to novel results for the shadow of Konoplya-Zhidenko naked singularity.

The paper is organized as follows. In Section II, we briefly introduce the spacetime of Konoplya-Zhidenko naked singularity, and study the unstable and stable, retrograde and prograde LRs. In Section III, we present the ring-shaped and arc-shaped shadows cast by the Konoplya-Zhidenko naked singularity, and investigate the influence of the naked singularity parameters and the observer's inclination angle on its shadow. Finally, we present a conclusion. In this paper, we employ the geometric units $G=c=M=1$.

\section{The spacetime of Konoplya-Zhidenko naked singularity and light rings}
\begin{figure}[htb]
\center{\includegraphics[width=16cm ]{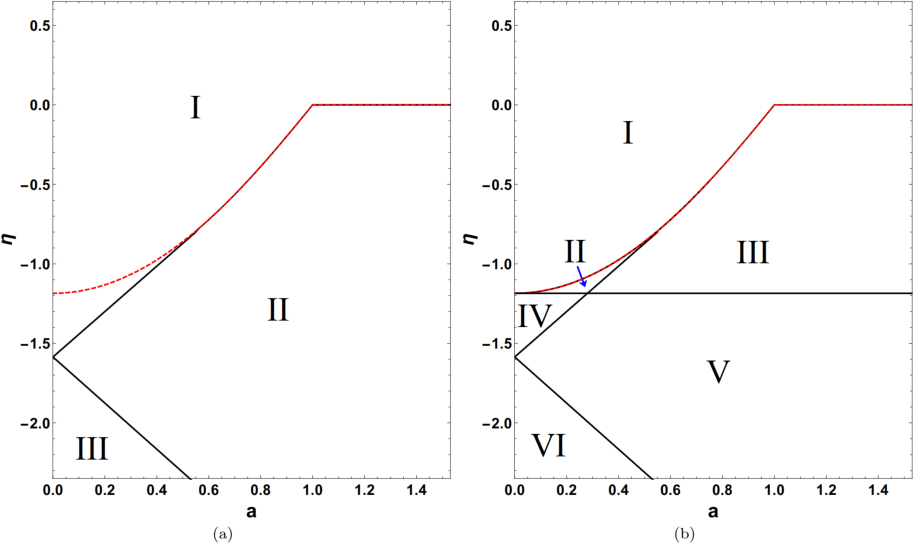}
\caption{The region of existence of the event horizon is the area above the red dashed curve. (a) The dependence of the existence of unstable LRs on the rotation parameter $a$ and deformation parameter $\eta$. In Region I, both the prograde and retrograde unstable LRs exist. In Region II, only the unstable retrograde LR exists. In Region III, neither the prograde nor retrograde unstable LR exists. (b) The dependence of the existence of stable LRs on $a$ and $\eta$. In Region I and VI, neither the prograde nor retrograde stable LR exists. In Region II, there are two stable prograde LRs. In Region III, there is one stable prograde LR. In Region IV, both the prograde and retrograde stable LRs exist. In Region V, only one stable retrograde LR exists.}
\label{fl}}
\end{figure}

Konoplya-Zhidenko spacetime describes an asymptotically flat, stationary, and axisymmetric spacetime with the deviation from the Kerr one through adding an extra deformation\cite{kz}. The metric of Konoplya-Zhidenko spacetime is described as follows:
\begin{eqnarray}
\label{xy}
ds^{2} &=& -(1-\frac{2Mr+\frac{\eta}{r}}{\rho^{2}})dt^{2}+\frac{\rho^{2}}{\Delta}dr^{2}+\rho^{2} d\theta^{2}+\sin^{2}\theta\bigg[r^{2}+a^{2}+\frac{2(M+\frac{\eta}{2r^{2}})ra^{2}\sin^{2}\theta}{\rho^{2}}\bigg]d\phi^{2}\\ \nonumber
&-&\frac{4(M+\frac{\eta}{2r^{2}})ra\sin^{2}\theta}{\rho^{2}}dtd\phi,
\end{eqnarray}
where
\begin{equation}
\Delta=a^{2}+r^{2}-2Mr-\frac{\eta}{r},\;\;\;\;\;\;\;\;\;\; \rho^{2}=r^{2}+a^{2}\cos^{2}\theta.
\end{equation}
$M$ is the mass of Konoplya-Zhidenko compact object, $a$ is the rotation parameter, and $\eta$ is the deformation parameter describing the deformation from the Kerr spacetime. When $\eta=0$, the metric will reduce to the usual Kerr metric. The position of event horizon can be defined by $\Delta=0$\cite{sy,sb10,suprr,quasin}. The condition for the existence of event horizon is
\begin{align}
\label{hcz}
\eta & >-\frac{2}{27}(\sqrt{4M^{2}-3a^{2}}+2M)^{2}(\sqrt{4M^{2}-3a^{2}}-M),&for \;\;\;|a|< M, \\
\eta & >0,&for \;\;\;|a|\geq M,
\end{align}
On the $(a, \eta)$ plane, the region of existence of the event horizon is the area above the red dashed curve in Fig.\ref{fl}. Comparing with the Kerr black hole, the deformation parameter $\eta$ extends the allowed range of the rotation parameter $a$. It allows $|a|\geq M$ for $\eta>0$.

The Hamiltonian of a photon propagation in Konoplya-Zhidenko spacetime can be characterized by
\begin{equation}
\label{hamil}
H(x,p)=\frac{1}{2}g^{\mu\nu}(x)p_{\mu}p_{\nu}=\frac{1}{2\rho^{2}}(p_{\theta}^{2}+\Delta p_{r}^{2}+V_{eff})=0,
\end{equation}
where the effective potential $V_{eff}$ is defined as
\begin{equation}
\label{veff}
V_{eff}=-\frac{1}{\Delta}[a L_{z}-(r^{2}+a^{2})E]^{2}+(\frac{L_{z}}{\sin\theta}-aE\sin\theta)^{2}.
\end{equation}
The energy $E$ and the $z$-component of the angular momentum $L_{z}$ of photon are two conserved quantities with the following forms
\begin{eqnarray}
\label{EL}
E=-p_{t}=-g_{tt}\dot{t}-g_{t\phi}\dot{\phi},\;\;\;\;\;\;\;\;\;\;\;\;\;\;
L_{z}=p_{\phi}=g_{\phi\phi}\dot{\phi}+g_{\phi t}\dot{t}.
\end{eqnarray}
The variables $r$ and $\theta$ in the Hamiltonian (\ref{hamil}) can be separated, so the null geodesic equations can be written as:
\begin{eqnarray}
\label{fc1}
R(r)&=&\Delta^{2} p_{r}^{2}=[aL_{z}-(r^{2}+a^{2})E]^{2}-\Delta K,\\
\Theta&=&p_{\theta}^{2}=K-\frac{1}{\sin^{2}\theta}(L_{z}-aE\sin^{2}\theta)^{2},
\label{fc}
\end{eqnarray}
where the quantity $K$ is the constant of separation associating with the hidden symmetries of the spacetime.

The PS is closely associated with the shadows of compact objects. It satisfies $\dot{r}=0$ and $\ddot{r}=0$, which is equivalent to
\begin{eqnarray}
\label{r}
R(r)&=&[aL_{z}-(r^{2}+a^{2})E]^{2}-\Delta K=0,\\ \nonumber
R'(r)&=&4Er[(r^{2}+a^{2})E-aL_{z}]-2K(r-M+\frac{\eta}{2r^{2}})=0.
\end{eqnarray}
Introducing two conserved parameters, the impact parameter $\xi$ and $\sigma$, as
\begin{eqnarray}
\label{xs1}
\xi=\frac{L_{z}}{E},\;\;\;\;\;\;\;\;\;\;\;\;\;\;
\sigma=\frac{K}{E^{2}}.
\end{eqnarray}
Solving the equations (\ref{r}), the constants $\xi$ and $\sigma$ of the spherical photon motion have the form,
\begin{eqnarray}
\label{pj}
\xi&=&\frac{a^2 \left(\eta-2M r^{2}-2r^{3}\right)+6 M r^4+5 \eta r^2-2 r^5}{a \left(2 r^3 -2M r^2+\eta\right)},\\
\sigma&=&\frac{16 r^5 \left(r^{3}+a^{2}r-2Mr^{2}-\eta\right)}{\left(2 r^3 -2M r^2+\eta\right)^2}.
\label{pj1}
\end{eqnarray}
$R''(r)>0$ represents the unstable photon spherical orbit (UPSO); $R''(r)<0$ represents the stable photon spherical orbit (SPSO).
$\Theta$ must be non-negative is deduced from Eq.(\ref{fc}). By substituting the constants $\xi$ (\ref{pj}) and $\sigma$ (\ref{pj1}) of the spherical photon motion into Eq.(\ref{fc}), one can obtain
\begin{eqnarray}
\label{ths}
\Theta_{s}&=&\sigma-\frac{1}{\sin^{2}\theta}(\xi-a\sin^{2}\theta)^{2}\\ \nonumber
&=&\frac{16 r^5 \left(r^{3}+a^{2}r-2Mr^{2}-\eta\right)}{\left(2 r^3 -2M r^2+\eta\right)^2}-\frac{1}{\sin^{2}\theta}\bigg[\frac{a^2 \left(\eta-2M r^{2}-2r^{3}\right)+6 M r^4+5 \eta r^2-2 r^5}{a \left(2 r^3 -2M r^2+\eta\right)}-a\sin^{2}\theta\bigg]^{2}.
\end{eqnarray}
The condition $\Theta_{s}\geq 0$ gives the region of the PS. $\Theta_{s}=0$ denotes the boundary of the PS, which consequently defines the silhouette of the black hole shadow. LR is given by
\begin{eqnarray}
\label{ghth}
\Theta_{s}|_{\theta=\frac{\pi}{2}}=\frac{16 r^5 \left(r^{3}+a^{2}r-2Mr^{2}-\eta\right)}{\left(2 r^3 -2M r^2+\eta\right)^2}-\bigg[\frac{a^2 \left(\eta-2M r^{2}-2r^{3}\right)+6 M r^4+5 \eta r^2-2 r^5}{a \left(2 r^3 -2M r^2+\eta\right)}-a\bigg]^{2}=0.
\end{eqnarray}
The radius $r_{c}$ of LRs can be determined by Eq.(\ref{ghth}), and its value depends on the rotation parameter $a$ and deformation parameter $\eta$. For rotating black holes, there are typically two LRs with different radii: one rotating in the same direction as the compact object's rotation, and the other rotating in the opposite direction. The radius $r_{c}$ of unstable prograde LR is smaller, with a positive impact parameter $\xi$, corresponding to the leftmost point of PS and black hole shadow. The unstable retrograde LR, on the other hand, has a larger $r_{c}$ and a negative $\xi$, corresponding to the rightmost point of PS and black hole shadow. So, the unstable prograde and retrograde LRs are crucial for the existence of a complete unstable PS and black hole shadow.

Figure \ref{fl}(a) exhibits the dependence of the existence of unstable LR on the rotation parameter $a$ and deformation parameter $\eta$. In Fig.\ref{fl}(a), the $(a, \eta)$ plane is divided into three regions by black curves: Region I, II and III. In Region I, both the prograde and retrograde unstable LRs exist; In Region II, only the unstable retrograde LR exists; In Region III, neither the prograde nor retrograde unstable LR exists. Due to the fact that the prograde and retrograde LRs are the leftmost and rightmost orbits within the PS, they are essential for a complete PS. Through our research, we found that the complete unstable PS exists only in Region I, while the unstable PSs in other regions are incomplete. When the event horizon exists in the Konoplya-Zhidenko spacetime (above the red dashed curve in Region I), both the prograde and retrograde unstable LRs exist, so does the complete unstable PS. However, in Region I, there is an area (below the red dashed curve in Region I) where the event horizon doesn't exist, yet the complete unstable PS could still exist. Figure \ref{fl}(b) exhibits the dependence of the existence of stable LRs on $a$ and $\eta$. In Fig.\ref{fl}(b), there are six regions I-VI divided by black curves. In Region I and VI, neither the prograde nor retrograde stable LR exists, which indicates the stable LR can not exist for Konoplya-Zhidenko black hole. There are only two stable prograde LRs in Region II and one stable prograde LR in Region III. In Region IV, both the prograde and retrograde stable LRs exist. In Region V, only one stable retrograde LR exists. For stable PS, we found that the complete stable PS exists only in Region IV, while the stable PSs in other regions are incomplete.

\section{The shadows of Konoplya-Zhidenko naked singularity}
In order to calculate the shadow of a compact object, celestial coordinates must be established in the observer's sky. In Refs.\cite{lf,sb10,my,sMN,mbw,mgw,scc,pe,halo,review}, we calculate the celestial coordinates in an axially symmetric spacetime as
\begin{eqnarray}
\label{ccxd}
&&x=-r\frac{p^{\hat{\phi}}}{p^{\hat{r}}}|_{(r_{o},\theta_{o})}, \nonumber\\
&&y=r\frac{p^{\hat{\theta}}}{p^{\hat{r}}}|_{(r_{o},\theta_{o})},
\end{eqnarray}
where $p^{\hat{\mu}}$ denotes the four-momentum of photons measured locally by the observer at ($r_{o}, \theta_{o}$). The locally measured four-momentum $p^{\hat{\mu}}$ can be expanded using the four-momentum $p^{\mu}$ of a photon as follows: \cite{lf,sb10,pe,halo,review,sha2,sw,swo,astro,chaotic,my,sMN,swo7,mbw,mgw,scc,dkerr}
\begin{eqnarray}
\label{kmbh}
p^{\hat{t}}&=&\sqrt{\frac{g_{\phi \phi}}{g_{t\phi}^{2}-g_{tt}g_{\phi \phi}}} E-\frac{g_{t\phi}}{g_{\phi\phi}}\sqrt{\frac{g_{\phi \phi}}{g_{t\phi}^{2}-g_{tt}g_{\phi \phi}}}L_{z}, \nonumber\\
p^{\hat{r}}&=&\frac{1}{\sqrt{g_{rr}}}p_{r},\;\;\;\;\;
p^{\hat{\theta}}=\frac{1}{\sqrt{g_{\theta\theta}}}p_{\theta},\;\;\;\;\;
p^{\hat{\phi}}=\frac{1}{\sqrt{g_{\phi\phi}}}L_{z},
\end{eqnarray}

The image points in the boundary of black hole shadow correspond to the light rays that spiral asymptotically toward the unstable PS. By substituting the constants $\xi$ and $\sigma$ of the PS (\ref{pj}, \ref{pj1}) into the celestial coordinates (\ref{ccxd}) and taking the limit as $r_{o}$ approaches $\infty$, one can derive the analytic expressions for the silhouette of Konoplya-Zhidenko compact object shadow\cite{lf,sb10,pe,halo,review,sha2,sw,swo,astro,chaotic,my,sMN,swo7,mbw,mgw,scc,dkerr},
\begin{eqnarray}
\label{xd1w}
x&=&-\frac{\xi}{\sin \theta_{o}}, \nonumber\\
y&=&\pm\sqrt{\sigma+2a\xi-\xi^{2}\csc^{2}\theta_{o}-a^{2}\sin^{2}\theta_{o}}.
\end{eqnarray}

\begin{figure}[htb]
\center{\includegraphics[width=16cm ]{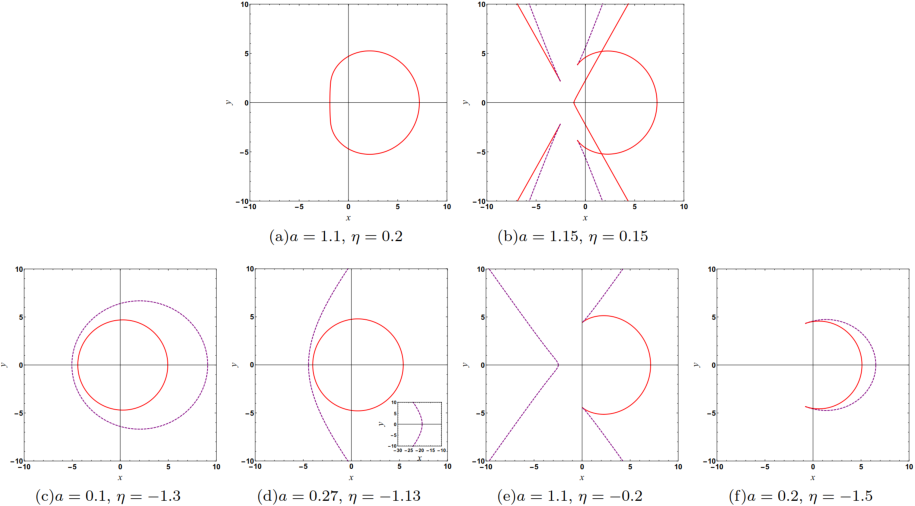}
\caption{Figures (a) and (b) represent the shadows (the area enclosed by the red lines) of the Konoplya-Zhidenko black hole, while Figures (c), (d), (e), and (f) depict the shadows (the red lines) of the Konoplya-Zhidenko naked singularity. The red lines represent the unstable PSs composed by UPSOs, and the purple dashed lines represent the stable PSs composed by SPSOs.}
\label{ggsd}}
\end{figure}

\begin{figure}[htbp]
\center{\includegraphics[width=8cm ]{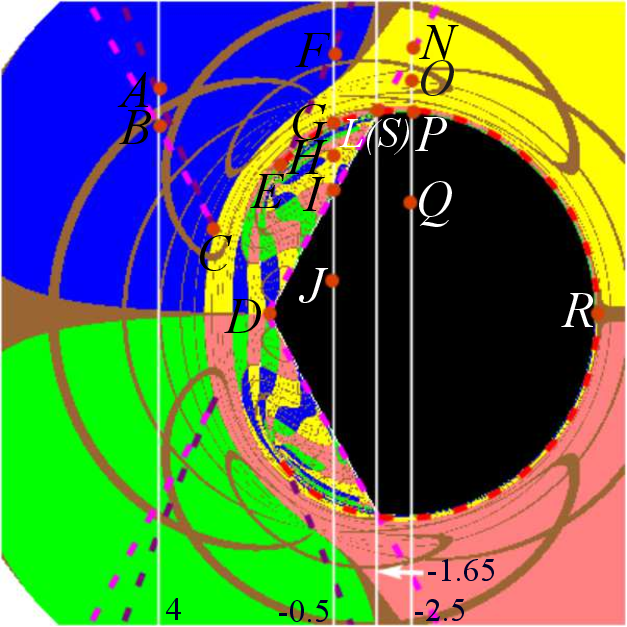}
\caption{The cusp shadow for the Konoplya-Zhidenko black hole with $a=1.15$ and $\eta=0.15$. The red and magenta dashed curves represent UPCOs; the purple dashed curve represents SPCOs.}
\label{1.15=0.15yy}}
\end{figure}

\begin{figure}[htb]
\center{\includegraphics[width=16cm ]{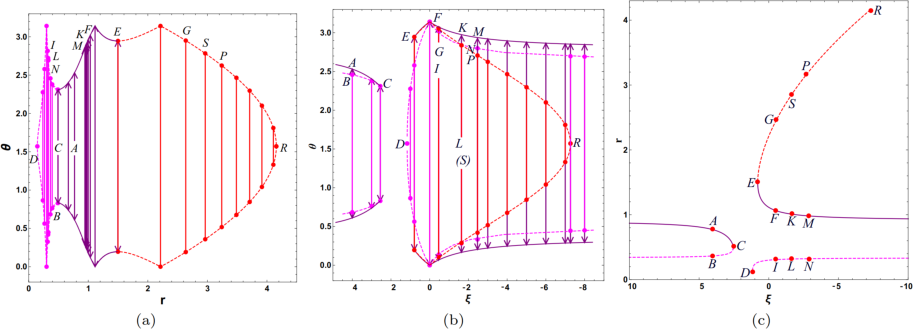}
\caption{The continua of UPCOs and SPCOs for $a=1.15$ and $\eta=0.15$ on the ($r, \theta$) plane, ($\xi, \theta$) plane, and ($r, \xi$) plane.}
\label{1.15=0.15etr}}
\end{figure}

\begin{figure}[htb]
\center{\includegraphics[width=12cm ]{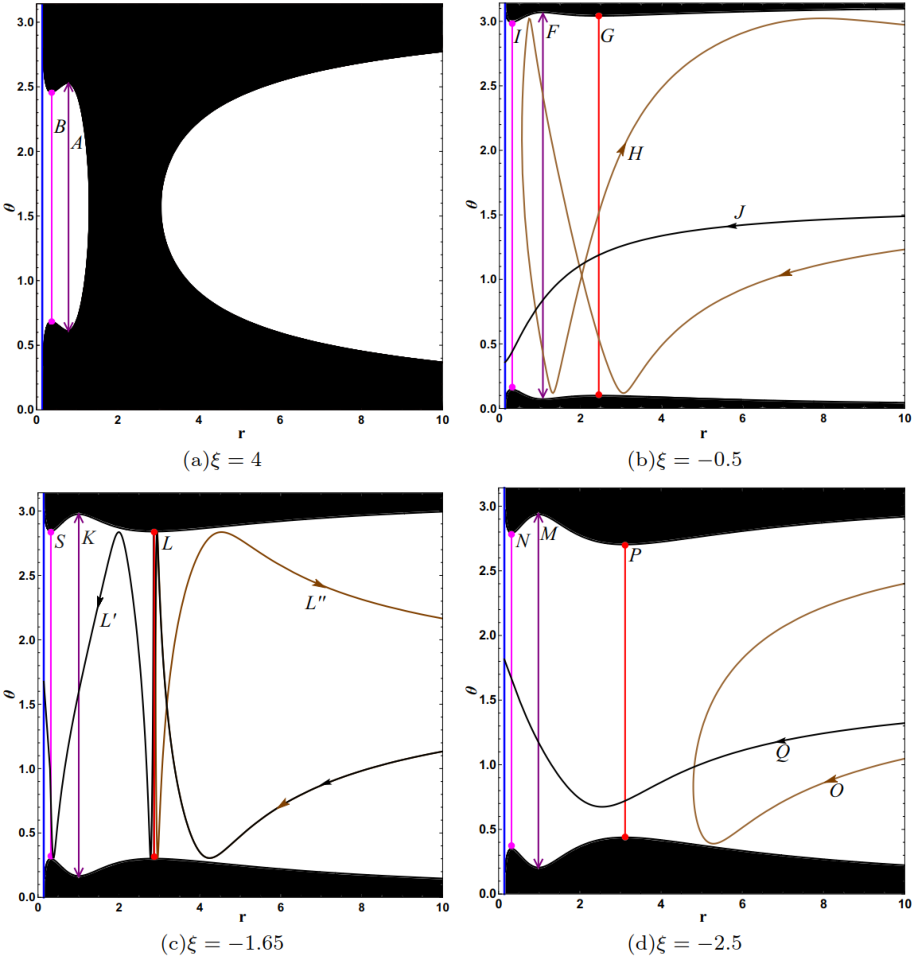}
\caption{The photon trajectories corresponding to the pixels $A, B, F, G, H, I, J, K, L, M, N, O, P, Q, R, S$ in Fig.\ref{1.15=0.15yy}. The blue line represents the event horizon of Konoplya-Zhidenko black hole, and the dark area represents the forbid region for photon($V_{eff}>0$).}
\label{1.15=0.15A}}
\end{figure}

\begin{figure}[htb]
\center{\includegraphics[width=8cm ]{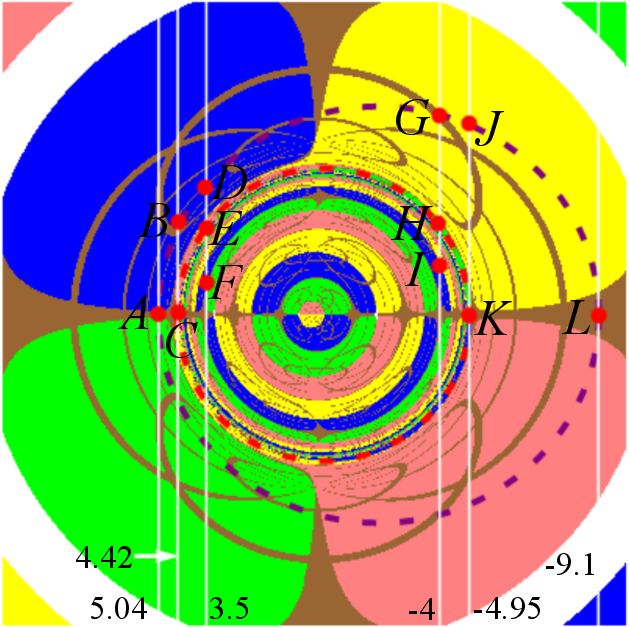}
\caption{The image of Konoplya-Zhidenko naked singularity with $a=0.1$ and $\eta=-1.3$. The red dashed circle represents the shadow (the complete unstable PS composed by UPSOs); the purple dashed circle represents the image of the complete stable PS composed by SPSOs.}
\label{0.1=-1.3yy}}
\end{figure}

Figure \ref{ggsd} displays the shadows of Konoplya-Zhidenko compact object under various scenarios depicted in Fig.\ref{fl}. The red lines represent the unstable PSs composed by UPSOs, and the purple dashed lines represent the stable PSs composed by SPSOs, which are plotted using analytical methods based on Eq.(\ref{xd1w}). In Fig.\ref{ggsd}(a) and (b), $a=1.1$, $\eta=0.2$ and $a=1.15$, $\eta=0.15$, respectively. In these cases, both event horizon and unstable PS exist, meaning that light rays entering the unstable PS will eventually reach the event horizon. Therefore, the shadow of Konoplya-Zhidenko black hole corresponds to the area enclosed by the unstable PS (the red circle). Figure \ref{ggsd}(b) displays three branches of UPSOs and two branches of SPSOs. However, it is only a subset of the UPSOs that define the shadow's boundary, leading to the distinctive cusp-shaped shadow of Konoplya-Zhidenko black hole, as we have researched in Ref.\cite{sb10}. This phenomenon is further illustrated in Fig.\ref{1.15=0.15yy}, \ref{1.15=0.15etr}, and \ref{1.15=0.15A}, with a detailed discussion. Figure \ref{ggsd}(c), (d), (e) and (f) exhibit the shadows of Konoplya-Zhidenko naked singularity. In Fig.\ref{ggsd}(c) and (d), $a=0.1$, $\eta=-1.3$ and $a=0.27$, $\eta=-1.13$, respectively. Differing from the Kerr naked singularity, in these scenarios, there not only exists a unstable retrograde LR but also a unstable prograde LR, enabling the formation of a complete unstable PS. But due to the absence of event horizon, the area enclosed by the unstable PS is not dark shadow. The shadow of Konoplya-Zhidenko naked singularity is only the image of unstable PS, shown as an ring-shaped shadow (the red circle). Furthermore, we show this ring-shaped shadow of Konoplya-Zhidenko naked singularity in Fig.\ref{0.1=-1.3yy} by the backward ray-tracing method\cite{lf,sb10,pe,halo,review,sha2,sw,swo,astro,chaotic,my,sMN,swo7,mbw,mgw,scc,dkerr}. In the cases of $a=1.1$, $\eta=-0.2$ (shown in Fig.\ref{ggsd}(e)) and $a=0.2$, $\eta=-1.5$ (shown in Fig.\ref{ggsd}(f)), the absence of unstable prograde LR results in an incomplete unstable PS, leading to the formation of an arc-shaped shadow (the red curve). In the Konoplya-Zhidenko black hole spacetime, the stable LR is absent. However, SPSOs may appear, as seen in the case of $a=1.15$ and $\eta=0.15$ (Fig.\ref{ggsd}(b)). But in the Konoplya-Zhidenko naked singularity spacetime, the stable LR could exist. For the case of $a=0.1$ and $\eta=-1.3$ (shown in Fig.\ref{ggsd}(c)), both prograde and retrograde stable LRs exist, resulting in the presence of a complete stable PS. The purple dashed ellipse in Fig.\ref{ggsd}(c) represents the image of the complete stable PS. As $a$ increases, we find the ellipse becomes more elongated and shifts to the right due to the drag effect. But due to the stability of the photon orbits in stable PS, observers cannot perceive the image of the stable PS. Therefore, we did not further investigate the stable PS. When $a=0.27$ and $\eta=-1.13$ (shown in Fig.\ref{ggsd}(d)), there are two stable prograde LRs and two branches of SPSOs. For the case of $a=1.1$ and $\eta=-0.2$ (shown in Fig.\ref{ggsd}(e)), one stable prograde LR and three branches of SPSOs emerge. When $a=0.2$ and $\eta=-1.5$ (shown in Fig.\ref{ggsd}(f)), only one stable retrograde LR and one branch of SPSO exist.

Figure \ref{1.15=0.15yy} shows the cusp shadow for the Konoplya-Zhidenko black hole with $a=1.15$ and $\eta=0.15$ with the backward ray-tracing method\cite{lf,sb10,pe,halo,review,sha2,sw,swo,astro,chaotic,my,sMN,swo7,mbw,mgw,scc,dkerr}. In the backward ray-tracing method, the light rays are assumed to evolve backward in time from the observer by solving the null geodesic equations numerically. Here we set the observer at $r_{o}=50, \theta_{o}=\pi/2$. The spherical background light source is the same as we set in Ref.\cite{my,sMN,mbw,mgw,scc,halo}, with a radius of $r_{s}=50$. In Fig.\ref{1.15=0.15yy}, the black area is Konoplya-Zhidenko black hole shadow, and the bright area is the image of the background light source. The red and magenta dashed curves represent UPCOs; the purple dashed curve represents SPCOs, which are plotted using Eqs.(\ref{xd1w}). The UPCOs can form a complete unstable PS to determine the boundary of black hole shadow. It can be observed that the shadows obtained through numerical methods and analytical methods are consistent. Figure \ref{1.15=0.15etr}(a), (b), and (c) show these UPCOs and SPCOs on the ($r, \theta$) plane, ($\xi, \theta$) plane, and ($r, \xi$) plane, respectively. $D$ represents the unstable prograde LR, giving rise to a continuum of UPCOs marked with $D-I-L-N$. $R$ represents the unstable retrograde LR, giving rise to a continuum of UPCOs $R-P-S-G-E$. Moreover, there is another continuum of UPCOs $C-B$ and two continua of SPCOs $C-A$ and $E-F$. These UPCOs and SPCOs, labeled with capital letters, are also marked in the black hole image (Fig.\ref{1.15=0.15yy}). It is shown that only the continua of UPCOs $D-I-L$ and $R-P-S$ determine the boundary of black hole shadow. These UPCOs can form a complete unstable PS $D-I-L(S)-P-R$. One can find that each UPSO or SPSO has an invariant radial coordinate $r$ and oscillating angular coordinate $\theta$ with respect to $\pi/2$. These photon trajectories are shown in Fig.\ref{1.15=0.15A}. Figure \ref{1.15=0.15A}(a) shows the SPCO $A$ and UPCO $B$. SPCO $A$ is stable, so it does not enter the black hole nor escape to infinity. Although UPCO $B$ can enter the black hole, it is isolated from the distant observers, so it is unrelated to the black hole shadow observed by observers. In Fig.\ref{1.15=0.15A}(b), one can find the UPCOs $G, I$, and SPCO $F$. However, only UPCO $I$ determines the boundary of the shadow, as it has a smaller passing angle $\Delta\theta$, as illustrated in Fig.\ref{1.15=0.15A}(b). The light rays that can pass through UPCO $G$ but not through UPCO $I$ still being unable to become part of the shadow, such as the light ray $H$ in Fig.\ref{1.15=0.15A}(b). The light ray $J$ passes through UPCO $I$, and enters the black hole, becoming part of the shadow. The lights $H$ and $J$ also are marked in Fig.\ref{1.15=0.15yy}. In Fig.\ref{1.15=0.15A}(c), one can find the UPCOs $L, S$, and SPCO $K$. The UPCOs $L$ and $S$ have the same passing angle $\Delta\theta$, but the UPCO $L$ is on the outer layer (has bigger radial coordinate $r$). The light rays $L'$ and $L''$ both reach UPCO $L$, but the light ray $L'$ passes through UPCO $L$, becoming part of the shadow, and the light ray $L''$ doesn't pass through UPCO $L$, escaping to infinity. In Fig.\ref{1.15=0.15A}(d), one can find the UPCOs $P, N$, and SPCO $M$. Only UPCO $P$ determines the boundary of the shadow due to it determines a smaller passing angle $\Delta\theta$. The light ray $Q$ passes through UPCO $P$ and become part of the shadow. The light ray $O$ doesn't pass through UPCO $P$ and escape to infinity. The lights $Q$ and $O$ also are marked in Fig.\ref{1.15=0.15yy}.

\begin{figure}[htb]
\center{\includegraphics[width=16cm ]{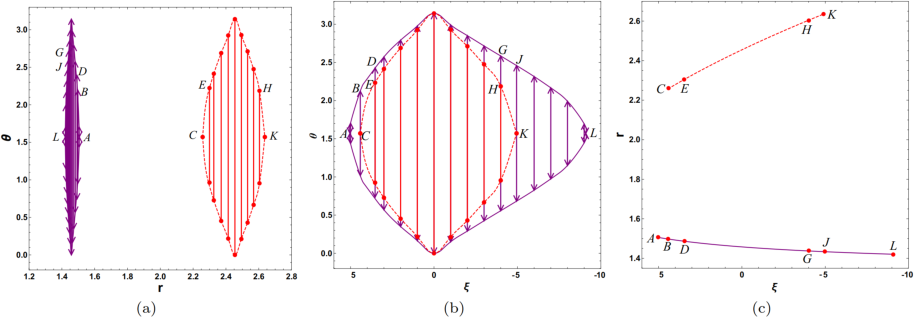}
\caption{The unstable PS composed by UPSOs and stable PS composed by SPSOs for $a=0.1$ and $\eta=-1.3$ on the ($r, \theta$) plane, ($\xi, \theta$) plane, and ($r, \xi$) plane.}
\label{0.1=-1.3etr}}
\end{figure}

\begin{figure}[htb]
\center{\includegraphics[width=16cm ]{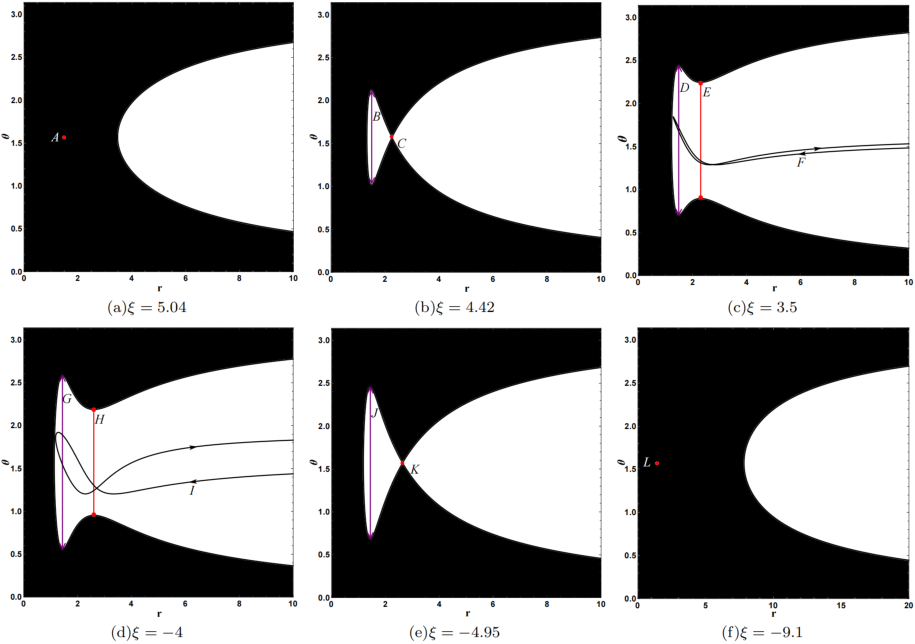}
\caption{The photon trajectories corresponding to the pixels $A, B, C, D, E, F, G, H, I, J, K, L$ in Fig.\ref{0.1=-1.3yy}. The dark area represents the forbid region for photon($V_{eff}>0$).}
\label{0.1=-1.3A}}
\end{figure}

By analytical methods, we have computed the images of complete PS for Konoplya-Zhidenko naked singularity, shown in Fig.\ref{ggsd}(c) and (d). Similar to the PS for black hole, it is also a circle. So, it is easy to mistakenly perceive the interior region of the PS as the shadow. Therefore, we plot the shadow of Konoplya-Zhidenko naked singularity with $a=0.1$, $\eta=-1.3$ by the backward ray-tracing method\cite{lf,sb10,pe,halo,review,sha2,sw,swo,astro,chaotic,my,sMN,swo7,mbw,mgw,scc,dkerr} in Fig.\ref{0.1=-1.3yy}. It can be observed that in Fig.\ref{0.1=-1.3yy}, no dark disk shadow is present; instead, there are an infinite number of relativistic Einstein rings\cite{can}, as indicated by the red dashed circle. It is, in fact, the image of the unstable PS, resulting from light rays spiraling toward the unstable PS from various directions and eventually reaching the observer. The purple dashed circle in Fig.\ref{0.1=-1.3yy} indicates the presence of a complete stable PS, which cannot be observed in the image of the Konoplya-Zhidenko naked singularity. Figure \ref{0.1=-1.3etr}(a), (b), and (c) show the complete unstable and stable PSs on the ($r, \theta$) plane, ($\xi, \theta$) plane, and ($r, \xi$) plane, respectively. The complete unstable PS is compose of a continuum of UPSOs marked with $C-E-H-K$, where $C$ and $K$ represent the unstable prograde and retrograde LRs. The complete stable PS is compose of a continuum of SPSOs marked with $A-B-D-G-J-L$, where $A$ and $L$ represent the stable prograde and retrograde LRs. These UPSOs and SPSOs, labeled with capital letters, are also marked in the image of Konoplya-Zhidenko naked singularity (Fig.\ref{0.1=-1.3yy}). One can find that SPSOs have smaller radii than UPSOs. Fig.\ref{0.1=-1.3A} exhibits the photon trajectories corresponding to the pixels $A, B, C, D, E, F, G, H, I, J, K$ and $L$ in Fig.\ref{0.1=-1.3yy}, where the dark area represents the forbid region for photon($V_{eff}>0$). In Fig.\ref{0.1=-1.3A}(a) and (f), $A$ and $L$ are the stable prograde and retrograde LRs with impact parameter $\xi=5.04$ and $-9.1$, respectively. In Fig.\ref{0.1=-1.3A}(b) and (e), $C$ and $K$ are the unstable prograde and retrograde LRs with impact parameter $\xi=4.42$ and $-4.95$, respectively. In Fig.\ref{0.1=-1.3A}(b-e), $B, D, G$ and $J$ represent SPSOs, and $E$ and $H$ represent UPSOs. In Fig.\ref{0.1=-1.3A}(c) and (d), the light rays $F$ and $I$ pass through UPSOs $E$ and $H$ respectively, meaning that they enter the PS, but then they turn back, escaping to infinity.

\begin{figure}[htb]
\center{\includegraphics[width=16cm ]{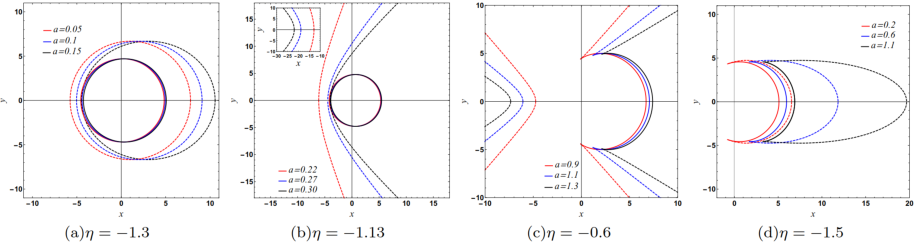}
\caption{The variations of the shadows of Konoplya-Zhidenko naked singularity with respect to spin parameter $a$ under various scenarios. The lines represent the shadows (the unstable PSs), and the dashed lines represent the images of the stable PSs composed by SPSOs.}
\label{sdbha}}
\end{figure}

\begin{figure}[htb]
\center{\includegraphics[width=16cm ]{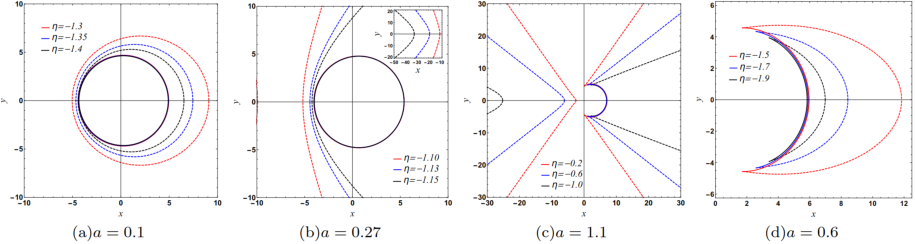}
\caption{The variations of the shadows of Konoplya-Zhidenko naked singularity with respect to deformation parameter $\eta$ under various scenarios. The lines represent the shadows (the unstable PSs), and the dashed lines represent the images of the stable PSs composed by SPSOs.}
\label{sdbhet}}
\end{figure}

Figure \ref{sdbha} illustrates the variations of the shadows of Konoplya-Zhidenko naked singularity with respect to spin parameter $a$ under various scenarios. The lines represent the shadows (the unstable PSs), and the dashed lines represent the images of the stable PSs composed by SPSOs. Regardless of whether it is a ring-shaped or arc-shaped shadow, they both shift gradually to the right as $a$ increases, similar to the shadow of Kerr black hole. In Fig.\ref{sdbha}(a) and (d), the image of the stable PS also shifts to the right as $a$ increases. In Fig.\ref{sdbha}(b) and (c), one branch of SPSOs shifts to the right, while the other shifts to the left as $a$ increases. Figure \ref{sdbhet} illustrates the variation of the shadows of Konoplya-Zhidenko naked singularity with respect to deformation parameter $\eta$ under various scenarios. In Fig.\ref{sdbhet}(a) and (b), the ring-shaped shadow exhibits almost no change with $\eta$. In Fig.\ref{sdbhet}(c) and (d), the arc-shaped shadow gradually becomes smaller as $|\eta|$ increases. The image of the stable PS becomes smaller as $|\eta|$ increases in Fig.\ref{sdbhet}(a) and (d). And one branch of SPSOs shifts to the right, while the other shifts to the left as $|\eta|$ increases in Fig.\ref{sdbhet}(b) and (c).

\begin{figure}[htb]
\center{\includegraphics[width=16cm ]{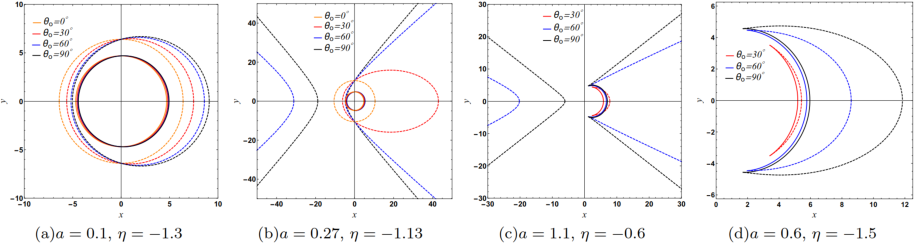}
\caption{The variations of the shadows of Konoplya-Zhidenko naked singularity with respect to the observer's inclination angle $\theta_{o}$ under various scenarios. The lines represent the shadows (the unstable PSs), and the dashed lines represent the images of the stable PSs composed by SPSOs.}
\label{sdjd}}
\end{figure}

Figure \ref{sdjd} illustrates the variation of the shadows of Konoplya-Zhidenko naked singularity with respect to the observer's inclination angle $\theta_{o}$ under various scenarios. The lines represent the shadows (the unstable PSs), and the dashed lines represent the images of the stable PSs composed by SPSOs. In Fig.\ref{sdjd}(a), both the ring-shaped shadow and the image of complete stable PS shift to the right as $\theta_{o}$ increases, and the latter become bigger. In Fig.\ref{sdjd}(b), the ring-shaped shadow also shifts to the right as $\theta_{o}$ increases, and the image of the complete stable PS becomes two branches of SPSOs as $\theta_{o}$ increases. In Fig.\ref{sdjd}(c) and (d), the arc-shaped shadow shifts to the right and enlarges in size as $\theta_{o}$ increases. However, no shadow emerges when $\theta_{o}=0$. The image of the incomplete stable PS transforms into three branches of SPSOs as $\theta_{o}$ increases in Fig.\ref{sdjd}(c), and becomes bigger as $\theta_{o}$ increases in Fig.\ref{sdjd}(d).

\section{Conclusion}
We research the shadows of Konoplya-Zhidenko naked singularity. In the spacetime of Konoplya-Zhidenko naked singularity, not only can unstable retrograde LR exist, but also unstable prograde LR, leading to the formation of a complete PS. Due to the absence of an event horizon, a dark disc-shaped shadow does not appear; instead, a ring-shaped shadow is observed. This is because the light rays passing through the unstable PS eventually escape to infinity. Furthermore, we calculate the image of the Konoplya-Zhidenko naked singularity using the backward ray-tracing method and find that the ring-shaped shadow appears as an infinite number of relativistic Einstein rings. For some parameter values (such as $a=1.1, \eta=-0.2$ or $a=0.2, \eta=-1.5$), only the unstable retrograde LR exists, resulting in an incomplete unstable PS and consequently giving rise to the arc-shaped shadow for Konoplya-Zhidenko naked singularity. The shadow of Konoplya-Zhidenko naked singularity gradually shifts to the right as the rotation parameter $a$ increases, and gradually becomes smaller as the deformation parameter $|\eta|$ increases. Moreover, the stable LRs and SPSOs can also exist in Konoplya-Zhidenko naked singularity spacetime, but they have no effect on the image of naked singularity. This study demonstrates that rotating naked singularity can exhibit not only an arc-shaped shadow but also a ring-shaped shadow.

\section{\bf Acknowledgments}

This work was supported by the National Natural Science Foundation of China under Grant No. 12105151, the Shandong Provincial Natural Science Foundation of China under Grant No. ZR2020QA080, and was partially supported by the National Natural Science Foundation of China under Grant No. 11875026, 11875025 and 12035005.

\end{document}